\documentclass[twocolumn,prl,aps,showpacs,superscriptaddress]{revtex4}

\usepackage{graphics}
\def\mevc2 {MeV/c$^2$}
\def\mevc {MeV/c}

\def\gevc2 {GeV/c$^2$}
\def\gevc {GeV/c}

\def\meandedx {$\langle$dE/dx$\rangle$}

\def\mtm {m$_t$-m$_0$}

\begin{document}
\bibliographystyle{prsty}

\title{Longitudinal Flow of Protons from 2-8 AGeV Central Au+Au Collisions}

\author{J.L.~Klay}	\affiliation{University of California, Davis, California 95616}
\author{N.N.~Ajitanand}	\affiliation{Department of Chemistry and Physics, SUNY, Stony Brook, New York 11794-3400}
\author{J.M.~Alexander}	\affiliation{Department of Chemistry and Physics, SUNY, Stony Brook, New York 11794-3400}
\author{M.G.~Anderson}	\affiliation{University of California, Davis, California 95616}
\author{D.~Best}	\affiliation{Lawrence Berkeley National Laboratory, Berkeley, California 94720}
\author{F.P.~Brady}	\affiliation{University of California, Davis, California 95616}
\author{T.~Case}	\affiliation{Lawrence Berkeley National Laboratory, Berkeley, California 94720}
\author{W.~Caskey}	\affiliation{University of California, Davis, California 95616}
\author{D.~Cebra}	\affiliation{University of California, Davis, California 95616}
\author{J.L.~Chance}	\affiliation{University of California, Davis, California 95616}
\author{P.~Chung}	\affiliation{Department of Chemistry and Physics, SUNY, Stony Brook, New York 11794-3400}
\author{B.~Cole}	\affiliation{Columbia University, New York, New York 10027}
\author{K.~Crowe}	\affiliation{Lawrence Berkeley National Laboratory, Berkeley, California 94720}
\author{A.C.~Das}	\affiliation{The Ohio State University, Columbus, Ohio 43210}
\author{J.E.~Draper}	\affiliation{University of California, Davis, California 95616}
\author{M.L.~Gilkes}	\affiliation{Purdue University, West Lafayette, Indiana 47907-1396}
\author{S.~Gushue}	\affiliation{Brookhaven National Laboratory, Upton, New York 11973}
\author{M.~Heffner}	\affiliation{University of California, Davis, California 95616}
\author{A.S.~Hirsch}	\affiliation{Purdue University, West Lafayette, Indiana 47907-1396}
\author{E.L.~Hjort}	\affiliation{Purdue University, West Lafayette, Indiana 47907-1396}
\author{L.~Huo}		\affiliation{Harbin Institute of Technology, Harbin, 150001 People's Republic of China}
\author{M.~Justice}	\affiliation{Kent State University, Kent, Ohio 44242}
\author{M.~Kaplan}	\affiliation{Carnegie Mellon University, Pittsburgh, Pennsylvania 15213}
\author{D.~Keane}	\affiliation{Kent State University, Kent, Ohio 44242}
\author{J.C.~Kintner}	\affiliation{St. Mary's College of California, Moraga, California 94575}
\author{D.~Krofcheck}	\affiliation{University of Auckland, Auckland, New Zealand}
\author{R.A.~Lacey}	\affiliation{Department of Chemistry and Physics, SUNY, Stony Brook, New York 11794-3400}
\author{J.~Lauret}	\affiliation{Department of Chemistry and Physics, SUNY, Stony Brook, New York 11794-3400}
\author{C.~Law}		\affiliation{Department of Chemistry and Physics, SUNY, Stony Brook, New York 11794-3400}
\author{M.A.~Lisa}	\affiliation{The Ohio State University, Columbus, Ohio 43210}
\author{H.~Liu}		\affiliation{Kent State University, Kent, Ohio 44242}
\author{Y.M.~Liu}	\affiliation{Harbin Institute of Technology, Harbin, 150001 People's Republic of China}
\author{R.~McGrath}	\affiliation{Department of Chemistry and Physics, SUNY, Stony Brook, New York 11794-3400}
\author{Z.~Milosevich}	\affiliation{Carnegie Mellon University, Pittsburgh, Pennsylvania 15213}
\author{G.~Odyniec}	\affiliation{Lawrence Berkeley National Laboratory, Berkeley, California 94720}
\author{D.L.~Olson}	\affiliation{Lawrence Berkeley National Laboratory, Berkeley, California 94720}
\author{S.Y.~Panitkin}	\affiliation{Kent State University, Kent, Ohio 44242}
\author{C.~Pinkenburg}	\affiliation{Department of Chemistry and Physics, SUNY, Stony Brook, New York 11794-3400}
\author{N.T.~Porile}	\affiliation{Purdue University, West Lafayette, Indiana 47907-1396}
\author{G.~Rai}		\affiliation{Lawrence Berkeley National Laboratory, Berkeley, California 94720}
\author{H.G.~Ritter}	\affiliation{Lawrence Berkeley National Laboratory, Berkeley, California 94720}
\author{J.L.~Romero}	\affiliation{University of California, Davis, California 95616}
\author{R.~Scharenberg}	\affiliation{Purdue University, West Lafayette, Indiana 47907-1396}
\author{L.~Schroeder}	\affiliation{Lawrence Berkeley National Laboratory, Berkeley, California 94720}
\author{B.~Srivastava}	\affiliation{Purdue University, West Lafayette, Indiana 47907-1396}
\author{N.T.B.~Stone}	\affiliation{Lawrence Berkeley National Laboratory, Berkeley, California 94720}
\author{T.J.M.~Symons}	\affiliation{Lawrence Berkeley National Laboratory, Berkeley, California 94720}
\author{S.~Wang}	\affiliation{Kent State University, Kent, Ohio 44242}
\author{R.~Wells}	\affiliation{The Ohio State University, Columbus, Ohio 43210}
\author{J.~Whitfield}	\affiliation{Carnegie Mellon University, Pittsburgh, Pennsylvania 15213}
\author{T.~Wienold}	\affiliation{Lawrence Berkeley National Laboratory, Berkeley, California 94720}
\author{R.~Witt}	\affiliation{Kent State University, Kent, Ohio 44242}
\author{L.~Wood}	\affiliation{University of California, Davis, California 95616}
\author{W.N.~Zhang}	\affiliation{Harbin Institute of Technology, Harbin, 150001 People's Republic of China}
\collaboration{E895 Collaboration}	\noaffiliation

\date{\today}

\begin{abstract}
Rapidity distributions of protons from central $^{197}$Au + $^{197}$Au collisions measured by the 
E895 Collaboration in the energy range from 2 to 8 AGeV at the Brookhaven AGS are presented.  Longitudinal 
flow parameters derived using a thermal model including collective longitudinal expansion are extracted from 
these distributions.  The results show an approximately linear increase in the longitudinal flow 
velocity, $\langle\beta\gamma\rangle_{L}$, as a function of the logarithm of beam energy.
\end{abstract}


\maketitle

                        
	Experimental heavy ion programs from the Bevalac, SIS, the AGS and the CERN
SPS have attempted to characterize the distributions of particles emerging from high energy collisions in terms of 
simple thermodynamic principles.  Testing the underlying assumptions of chemical and thermal equilibrations at 
different stages of the collision has been attempted by comparing model expectations to many different experimental observables.  Static 
isotropic thermal emission models applied to observed particle rapidity distributions from experiments at all beam energies consistently 
fail to describe the observed shape; such model predictions being too narrow.  Thermal models 
which include collective longitudinal expansion have been much more successful at reproducing the observed 
rapidity distributions \cite{Schn93,Stac96}.  The additional collective motion is attributed to intense pressure 
gradients which develop in the very hot, compressed nuclear matter fireballs created in heavy ion collisions.

	At the AGS, rapidity distributions of multiple particle species, including pions, kaons, protons and 
lambda hyperons from central collisions have been simultaneously described by a thermal distribution with 
a common longitudinal expansion velocity \cite{Stac96,Brau95}.  The agreement between the 
proton and other particle distributions suggests a high degree of stopping of the incident nucleons at 
the top AGS energy, which implies even more stopping at lower beam energies.  However, recent 
investigations of the centrality dependence of the proton rapidity distributions from 6-11 AGeV Au+Au 
collisions by E917 suggest that the degree of nucleon stopping may be less than previously 
considered \cite{Back01}.  Nevertheless, their flat dN/dy for {\it central} collisions at all beam 
energies, fitted by sources distributed uniformly in rapidity to y-y$_{cm}$ = 0, could also be interpreted in the 
manner presented herein.

	For the asymptotic case at extremely high beam energies, Bjorken proposed \cite{Bjor83} 
that nuclear transparency would evacuate the central rapidity region of all of the initial nucleons, leaving a 
hot, high-energy density region in which a Quark Gluon Plasma might form.  In 160 AGeV Pb+Pb collisions at the 
CERN SPS, observed net proton ((+)-(-)) rapidity distributions \cite{Appe99} exhibit a double-humped character 
which is consistent with this transparency.  At SPS energies, the nucleon distributions are not describable by a 
simple thermal model with only collective longitudinal flow, but rather need additional theoretical consideration of 
transparency.  However, the negative hadron (mostly pions) rapidity distributions are well-described by this model, 
which might be interpreted as suggesting a significant amount of collective longitudinal flow for produced 
particles \cite{Stac96}.

	More recently, models which attempt simultaneously to characterize many of the freeze-out parameters 
(transverse and longitudinal flow, temperature, chemical potential and source volume) have been attempted 
on data at and above the top energy at the AGS \cite{Dobl99}.  The data discussed in this letter are 
compared only to the locally thermal model with longitudinal flow.

	The new data were taken by the E895 Experiment \cite{E895} at nominally 2, 4, 6, and 8 AGeV 
(1.85, 3.91, 6.0 and 8.0 AGeV after correction for energy loss before the target) at the 
Brookhaven AGS using the EOS TPC \cite{Wiem91}.  Global characterization of the 
collisions is made possible by the nearly 4$\pi$ center of mass frame coverage provided by this detector.  Particle identification is 
achieved by correlating the average ionization energy loss in the P10 drift gas with the charged 
particle momenta, reconstructed from the curvature of tracks bent by the Multi-Particle
Spectrometer (MPS) magnet as they pass through the TPC.

	The 5\% most central events were selected using the distribution of primary charged particle event multiplicities.  Primary tracks 
are required to originate within 2.5 cm of the reconstructed event vertex. This analysis is based on approximately 20,000 selected 
central events at each beam energy.

For particle identification \cite{My Thesis}, all tracks are initially assigned the proton mass and sorted 
into rapidity and transverse mass bins: 0.1 unit rapidity slices from beam to 
target rapidity, with the mid-rapidity slice corresponding to $|y-y_{cm}|$ $<$ 0.05, and 25 MeV/c$^2$ wide 
transverse mass bins in the range 0 $<$ {\mtm} $<$ 1.0 GeV/c$^2$.  The mean total momentum and $\beta\gamma$ 
are computed at the center of the ({\mtm},y) bins and are used with a parameterization of the Bethe-Bloch
energy loss as a function of $\beta\gamma$ to fit for the relative particle populations in {\meandedx} projected for each 
({\mtm},y) bin.  The total yield of protons in each bin is obtained from the area under the fitted proton {\meandedx} distribution.

	Up to a total momentum, p, of 1 GeV/c, the protons are isolated in {\meandedx} and easily 
identified.  Above 1 GeV/c, pions and kaons begin to 
contaminate the proton sample.  Between 6 and 8 GeV/c total momentum, the proton, deuteron and triton {\meandedx} 
distributions significantly overlap each other and their individual yields cannot be resolved.  
Pion contamination is removed by using the observed ratio of oppositely charged pions in the momentum range up to 1 GeV/c to extrapolate 
to higher momenta \cite{My Thesis}.  The relatively cleanly observed negative pion yields, combined with the extrapolated pion ratios, are 
used to remove the $\pi^+$ contamination from the proton distributions above 1 GeV/c.  Kaon contamination is eliminated by using the 
measured kaon results from \cite{Ogil98,Dunl99} for the same beam energies.  The confusion among protons, deuterons and 
tritons between 6 and 8 GeV/c results in a broad hole in the final proton spectra, where the proton yields cannot be extracted.
Where appropriate, the reported errors for the raw proton yields 
include an estimate of the additional uncertainty resulting from the models used to determine the contamination from other particles.

\begin{figure*}
\resizebox*{0.7\textheight}{!}{\includegraphics{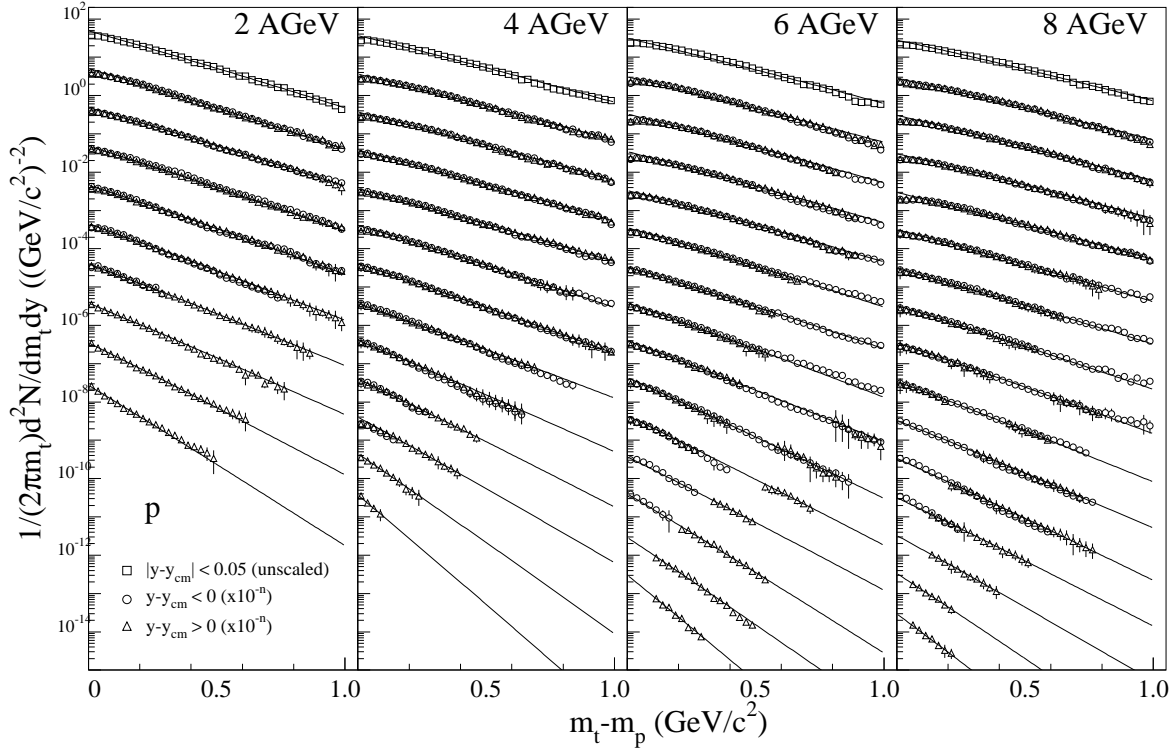}}
\caption{Invariant yield per event as a function of {\mtm} for protons in central Au+Au collisions at 2, 4, 6, and 8 AGeV.
Midrapidity is shown unscaled, while the 0.1 unit forward/backward rapidity slices are scaled down
by successive factors of 10.} 
\label{fig:specmt}
\end{figure*}

	Simulations of the detector response to protons in all regions of phase space were 
performed using GEANT 3.21.  Small samples ($\leq$ 4 tracks per event) of  
proton tracks with momentum distributions approximating the measured data were embedded into raw events and propagated through the 
reconstruction chain.  A correction for detector efficiency, acceptance and momentum resolution was obtained from the ratio of found GEANT 
tracks to the simulated input using the same track quality cuts and centrality selection as the data to 
be corrected.  The proton corrections were largest at very backward rapidities, where detector acceptance 
causes significant losses, and at forward rapidities, where track density is large, and at low transverse 
mass.  However, the corrected spectra, Fig.\ \ref{fig:specmt}, show very good forward/backward 
rapidity reflection symmetry.

	
	The transverse mass spectra of the protons in each rapidity slice are shown 
in Fig.\ \ref{fig:specmt}.  In order to determine yields, these spectra have been fit with the 
Maxwell-Boltzmann distribution.  
\begin{equation}
 { {1 \over 2\pi m_{t}} {d^{2}N \over dm_{t}dy} } = { A(y) m_{t} e^{-(m_{t}-m_{0})/T(y)} },
\label{eq:Thermal fit}
\end{equation}
where $A(y)m_{0}$ represents the {\mtm}=0 intercept and $T(y)$ is the inverse slope parameter.  
It agrees fairly well at large {\mtm}, but the agreement deteriorates 
for very low {\mtm}, where transverse flow modifies Eq. (\ref{eq:Thermal fit}).  However, the 
resulting systematic overestimate of the total integrated yields obtained from the fits are smaller than 
the uncertainties in the measured integrated yields, dN/dy.

        The rapidity density from Eq. (\ref{eq:Thermal fit}) can be obtained by integrating over $m_{t}$ 
with an overall normalization parameter, B, as
\begin{equation}
{dN_{th} \over dy}(y) = B T^{3} ({{m^2 \over T^2} + {m \over T} {2
\over \cosh y} + {2 \over \cosh^2 y}})
e^{({- {m \over T} \cosh y})}.
\label{eq:Thermal dNdy}
\end{equation}

	The width of the distribution, Eq. (\ref{eq:Thermal dNdy}), as a function of rapidity for a given
emitted particle is a function of the temperature and particle mass only.  Using the 
mid-rapidity inverse slope parameters (uncertainty of 2 MeV) from Eq. (\ref{eq:Thermal fit}) to approximate the temperature 
(Table \ref{Long Flow Table}) in Eq. (\ref{eq:Thermal dNdy}), this distribution can be compared to the rapidity 
densities extracted from the integrated transverse mass spectra.  The result, shown as dashed lines in Fig.\ \ref{Protons with 
long flow}, is clearly too narrow to describe our data.

	Schnedermann, et al.\ \cite{Schn93} modeled this increase in the widths as the result of collective 
longitudinal flow.  This assumes that the observed distributions arise from the superposition of multiple boosted 
individual isotropic, locally thermalized sources at each rapidity, $\eta$.  This is the longitudinally 
boost-invariant approach \cite{Bjor83} but taken to a finite boost, $\eta_{max}$.  
The distributions are 
\begin{equation}
\label{eq:longflow}
{dN \over dy} = {\intop}^{\eta_{max}}_{\eta_{min}} d\eta {dN_{th}
\over dy}{(y - \eta)}
\end{equation}
where $\eta_{max}$ = - $\eta_{min}$, from symmetry about the center of mass; $dN_{th}/dy$ is Eq. 
(\ref{eq:Thermal dNdy}) with T from Table \ref{Long Flow Table}; and $N$ is the number of protons 
observed.  An average longitudinal source velocity can be obtained by averaging over $|\eta|$ as 
$\langle\beta_{L}\rangle = tanh(\eta_{max}/2)$.

	Fits of this form have been applied to a wide array of experimental data at the AGS and 
the SPS \cite{Schn93,Stac96,Brau95,Herr99}.  Although the inverse slope in Table \ref{Long 
Flow Table} is not the true temperature, the longitudinal boost, $\eta_{max}$, has been observed to depend 
only weakly on the temperature input to the model, with the exception of nucleons at SPS energies, 
which are affected by transparency \cite{Schn93}.   In the present data, a 50\% decrease in the 
temperature results in a maximum change of only 15\% in $\eta_{max}$.

\begin{figure}
\resizebox*{0.5\textwidth}{!}{\includegraphics{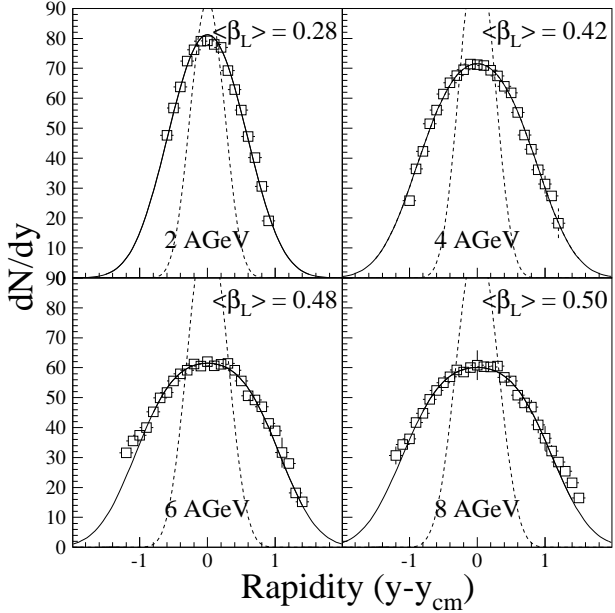}}
\caption{Proton rapidity distributions in central Au+Au collisions at 2, 4, 6, and 8 AGeV.  The
dashed curves correspond to isotropic emission from a
stationary thermal source with temperatures given by the mid-rapidity inverse slope
parameters from the transverse mass fits (Eq. (\ref{eq:Thermal dNdy})), whereas the solid curves 
indicate fits with longitudinal flow (Eq. (\ref{eq:longflow})).}
\label{Protons with long flow}
\end{figure}

	Fits of Eq. (\ref{eq:longflow}) to the proton rapidity densities at 2, 4, 6, and 8 AGeV are also 
shown on Fig.\ \ref{Protons with long flow}.  The fit parameter, $\eta_{max}$, and the corresponding average velocities, 
$\langle\beta_{L}\rangle$, are listed in Table \ref{Long Flow Table}.  $\langle\beta\gamma\rangle_{L}$ is 
evaluated by computing $\gamma = 1/\sqrt{1-\langle\beta_{L}\rangle^2}$.

\begin{table}
{\centering \begin{tabular}{cccccc}
E$_{beam}$ & T$_0$ & $\eta_{max}$ & $\langle\beta_{L}\rangle$ & 
$\langle\beta\gamma\rangle_{L}$ & Ref.\\
($AGeV$) & ($MeV/c^2$) & & & & \\
\hline
1.06 (Au)& 80 & 0.37 & 0.19 & 0.194 & \\
1.15 (Au)& 92 & 0.41 & 0.20 & 0.204 & {\it \cite{Liu98}} \\
2 & 187 & 0.570 $\pm$ 0.024 & 0.28 & 0.292 &\\
4 & 211 & 0.889 $\pm$ 0.024 & 0.42 & 0.463 &\\
6 & 216 & 1.049 $\pm$ 0.026 & 0.48 & 0.547 &\\
8 & 229 & 1.104 $\pm$ 0.025 & 0.50 & 0.577 &\\
6 (Au)& 253 & 0.990 & 0.46 & 0.518 & {\it \cite{Back01}} \\
8 (Au)& 267 & 1.086 & 0.50 & 0.577 & {\it \cite{Back01}} \\
10.8 (Au)& 279 & 1.166 & 0.52 & 0.609 & {\it \cite{Back01}} \\
11 (Au)& 130 & 1.10 & 0.50 & 0.577 & {\it \cite{Stac96}} \\
14.6 (Si+Al)& 120 & 1.15 & 0.52 & 0.609 & {\it \cite{Brau95}} \\
158 (Pb)& 160 & 1.99 & 0.76 & 1.169 & {\it \cite{Stac96}} \\
200 (S)& 220 & 1.70 $\pm$ 0.30 & 0.69 & 0.953 & {\it \cite{Schn93}} \\ 
\end{tabular}\par}
\caption{Longitudinal flow parameters extracted from fits of Eq. (\ref{eq:longflow}) to the 
proton rapidity densities at 2, 4, 6, and 8 AGeV, using the mid-rapidity proton inverse slope parameters, T$_0$.  
Note that the parameters extracted at the SPS come from produced particles, primarily pions, with 
comparison among other species, such as kaons and strange baryons suggesting collectivity.}
\label{Long Flow Table}
\end{table}

\begin{figure}
\resizebox*{0.5\textwidth}{!}{\includegraphics{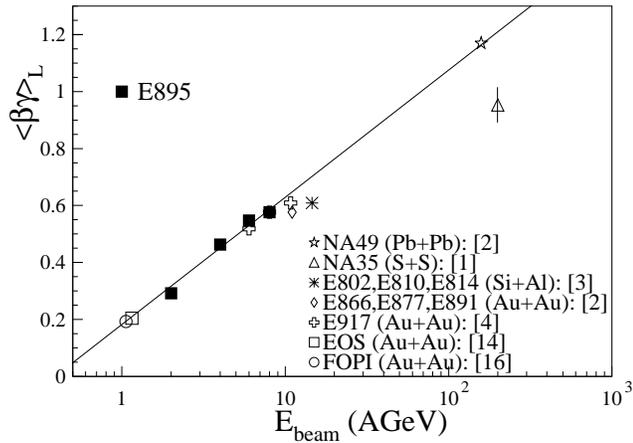}}
\caption{Energy excitation function of longitudinal flow velocities
($\langle\beta\gamma\rangle_{L}$), from heavy ion collisions.  The open circle has been adjusted here as 
described in the text.  For heavy systems, a roughly linear dependence with respect to $\log_{10}(E_{beam})$ over more than two 
orders of magnitude of beam energy is illustrated (solid line).  Note that the parameters extracted at the SPS come 
from produced particles, primarily pions, with comparison among other species, such as kaons and strange baryons 
suggesting collectivity.}
\label{World's long flow}
\end{figure}

	Fig.\ \ref{World's long flow} shows our results for $\langle\beta\gamma\rangle_{L}$ as a function of beam 
energy.  Also included are the extracted parameters from the compiled data from \cite{Schn93,Stac96,Brau95,Back01,Herr99,Liu98}.

	A modification to a previous interpretation of the lowest energy FOPI data point is presented here.  
The proton longitudinal flow reported for 1.06 AGeV Au+Au in \cite{Stac96,Herr99,Wess97} is from an 
interpretation of the measured data from \cite{Herr96} which assumed isotropic emission from a Siemens and  
Rasmussen \cite{Siem79} radially boosted spherical thermal source.  Good agreement between 
their Au+Au rapidity distributions and predictions from a Siemens-Rasmussen style model based on 
fits to their mid-rapidity kinetic energy spectra was reported \cite{Herr96}.  A temperature of 
80 MeV and an average (isotropic) flow velocity of $\langle\beta_{r}\rangle$ = 0.41 were reported.

	The resulting longitudinal flow velocity of $\langle\beta\rangle$ = 0.38 attributed to the 1.06 AGeV
Au+Au data by \cite{Stac96} is a factor of two larger than the value one 
extracts using the present prescription.   The value reported in Table \ref{Long Flow Table} and 
Fig.\ \ref{World's long flow} is from our re-analysis of these FOPI 1.06 AGeV rapidity densities using their 
reported temperature of 80 MeV with Eq. (\ref{eq:longflow}).  The result is $\langle\beta_{L}\rangle$ = 
0.19, which is consistent with a value obtained using the same model with Au+Au 
collision data measured by the EOS collaboration for E$_{beam}$=1.15 AGeV and reported in \cite{Liu98}.

	Based on the reported $\langle\beta_{L}\rangle$ for Au+Au at 1 AGeV and 11 AGeV, a linear relation between 
$\langle\beta_{L}\rangle$ and the logarithm of beam energy was suggested \cite{Stac96,Herr99,Wess97}.  
Extrapolated to SPS energies, it did not describe the longitudinal flow at the SPS 
extracted from produced particles such as pions.  However, with the inclusion of the present data from E895 and our 
re-analysis of the 1 AGeV protons, Fig.\ \ref{World's long flow} shows a linear trend, spanning 1 to 160 
AGeV, for the heaviest systems, with a slope $\sim$~2.5 larger than was 
previously concluded.  The SPS Pb+Pb value (obtained from produced particles, not protons) no longer shows 
a significant excess above the systematic.  
Further study of rapidity distributions for protons as a function of centrality \cite{Back01}, and produced 
particles, with system size and beam energy in the range above the AGS would help to clarify the relation 
between stopping and longitudinal expansion of protons in heavy ion collisions.


This work was supported in part by the US National Science Foundation under Grants No.
PHY-98-04672, PHY-9722653, PHY-96-05207, PHY-9601271, and PHY-9225096; by
the U.S. Department of Energy under grants DE-FG02-87ER40331.A008,
DE-FG02-89ER40531, DE-FG02-88ER40408, DE-FG02-87ER40324, and contracts
DE-AC03-76SF00098, DE-AC02-98CH10886; and by the University of Auckland
Research Committee, NZ/USA Cooperative Science Programme CSP 95/33; and by
the National Natural Science Foundation of P.R. China under
grant number 19875012.

\end{document}